# Solving the Coronal Heating Problem using X-ray Microcalorimeters

S. Christe[1], S. Bandler[1], E. DeLuca[2], A. Caspi[3], L. Golub[2], R. Smith[2], J. Allred[1], J. W. Brosius[4], B. Dennis[1], J. Klimchuk[1]

[1]NASA Goddard Space Flight Center, Greenbelt, MD, USA; [2]Harvard-Smithsonian Center for Astrophysics, Cambridge, MA, USA; [3]Southwest Research Institute, Boulder, CO, USA; [4]Catholic University, Washington, DC, USA

Even in the absence of resolved flares, the corona is heated to several million degrees. However, despite its importance for the structure, dynamics, and evolution of the solar atmosphere, the origin of this heating remains poorly understood. Several observational and theoretical considerations suggest that the heating is driven by small, impulsive energy bursts which could be Parker-style "nanoflares" (Parker 1988) that arise via reconnection within the tangled and twisted coronal magnetic field. The scale heights of coronal loops are clearly incompatible with static heating models (Petrie 2006). Recent models of the First Ionization Potential (FIP) effect where the ponderomotive force from Alfvén waves drives the fractionation (Laming 2009) strongly suggest the presence of impulsive events in the corona. The classical "smoking gun" (Klimchuk 2009; Cargill et al. 2013) for impulsive heating is the direct detection of widespread hot plasma (T > 6 MK) with a low emission measure.

Tantalizing hints of hot plasma have been seen from the solar corona imaged in the light of Mg XII 8.42 Å (Zhitnik et al. 2006; Urnov et al. 2007), in emission measure analyses of spectra taken with Hinode/EIS (Patsourakos & Klimchuk 2009), with Hinode/XRT using different filter ratios (Reale et al. 2009; Schmelz et al. 2009) and/or in quiet sun observations with RHESSI (McTiernan 2009; Reale et al. 2009b) as well as two rocket-borne instruments (Brosius et al. 2014; Caspi et al. 2015). Modeling suggests that nanoflares should be followed by episodes of evaporation from the chromosphere, following heat conduction downwards. This upflow is predicted to be 50 to 100 km s$^{-1}$. To date, such upflows have generally only been convincingly detected during solar flares (Brosius & Phillips 2004), when shifts of the entire line profile have been observed (Brosius 2003, 2013) or as a "shoulder" on the blue wing of an otherwise stationary line profile (Milligan et al. 2006).

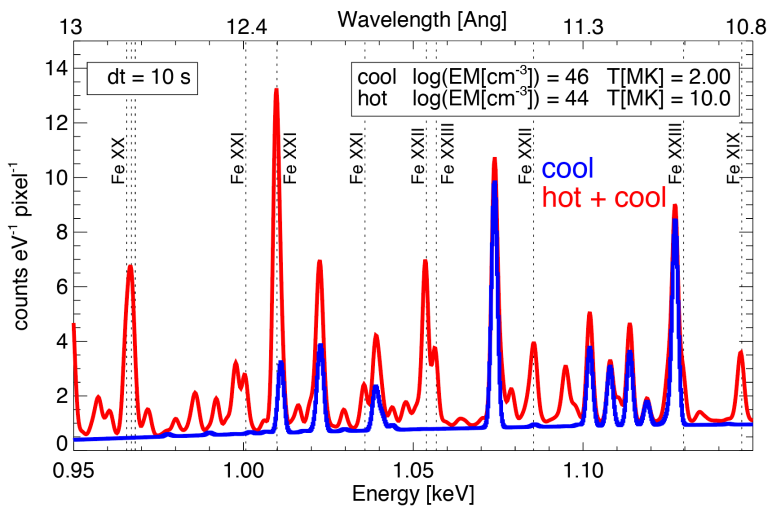

**Figure 1**. Simulated spectrum of an active region from a single TES pixel integrated over 10 seconds. The spectrum consists of two isothermal plasmas. The majority component is at 2 MK (blue, cool only). A hotter component at 10 MK has an emission measure that is 1% of the cooler component. The addition of the hot component (red) shows strong signatures from a number of additional highly ionized Fe lines.

These results have been obtained using a variety of technologies. Recent solar missions (Hinode, SOHO, IRIS) have used extreme ultraviolet (EUV) slit spectrographs to measure the velocity, temperature, and density in active regions and flares. EUV instruments have the advantage of very high spectral resolution across a wide range of emission lines. However, these devices have significant disadvantages. They require several minutes to scan (raster) across an active region. Furthermore since the exposure time determines the raster speed, selecting the correct exposure time is challenging for measuring temporally-varying features such as those found in active regions. The same can be said of Bragg Crystal Spectrometers. For EUV imagers which do not raster such as AIA on SDO, filter bandpasses are relatively broad compared to the underlying lines and frequently include low temperature emission which is difficult to isolate from the hot plasma. Additionally, EUV line spectroscopy has limited sensitivity to the very hottest temperatures, >20 MK, due to a lack of spectral lines from ion species formed at these temperatures.

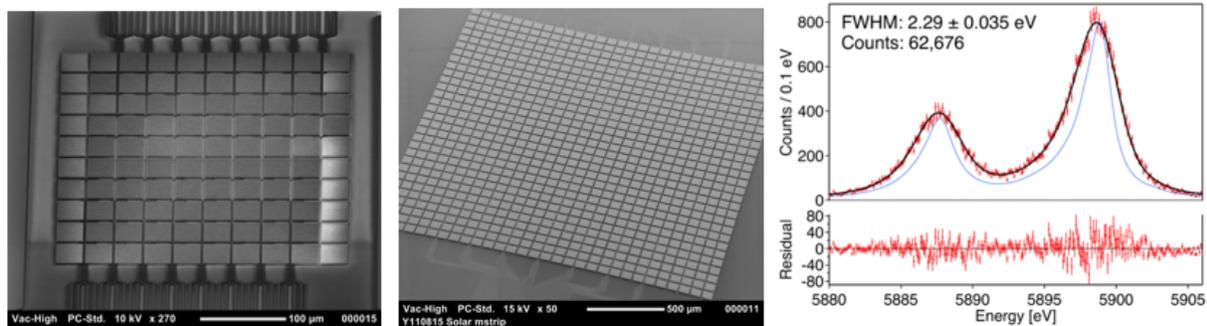

**Figure 2.** (left) Prototype TES X-ray microcalorimeter array with 35 µm pitch in which <2 eV energy resolution has been demonstrated (Datesman 2016). (middle) Prototype kilo-pixel microcalorimeter array with 75 µm pitch. (right) Spectrum of Mn Ka X-rays with a FWHM energy resolution <2.3 eV, with 99.6% throughput for an input count rate of 100 cps.

In recent years there has been great progress in the development of Transition Edge Sensor (TES) X-ray microcalorimeters that make them more ideal for studying the Sun. In this detector, a superconductor is biased at a temperature in-between its superconducting and normal-metal state, where it is extremely sensitive to small energy inputs such as absorption of X-ray photons. When combined with grazing-incidence focusing optics, they provide direct spectroscopic imaging over a broad energy band (0.5 to 10 keV), overcoming all of the issues described above. Close-packed arrays of pixels on pitches ranging from 35 to 75 µm have now been developed, as shown Fig. 2 (left, middle) allowing studies down to arcsecond angular scales (Bandler 2013; Datesman 2016). Extremely impressive energy resolution has been demonstrated in small pixels, as low as 0.7 eV (FWHM) at 1.5 keV (Lee 2015), and 1.56 eV (FWHM) at 6 keV (Smith 2012) -- two orders of magnitude better than the current best traditional solid state photon-counting spectrometers. Pixels have successfully operated at several hundred counts per second. Fig. 2 (right) shows a spectrum of Mn Ka X-rays collected at 100 counts per second, all around 6 keV, with an energy resolution of 2.3 eV FWHM. Kilo-pixel arrays have already been fabricated as shown in Fig. 2 (middle). Several abutted kilo-pixels arrays at the focal plane can provide an appropriate field of view. The technology needed for the multiplexed readout of large arrays has continued to develop and is now ready for flight. Most recently, code-division multiplexing demonstrated the multiplexed readout of 32 TESs without any significant energy resolution degradation (Morgan 2016). Progress has also been made in another promising readout technology called microwave SQUID multiplexing, which has the promise of being able to read out hundreds of pixels on a single readout channel (Bennett 2015).

The simultaneous combination of high resolution direct spectroscopic imaging observations across this entire energy range is new and tremendously powerful. The improved capabilities of a pixelated microcalorimeter, relative to EUV instruments, include:
- sensitivity to plasma temperatures from <0.7 MK to ~100 MK
- the ability to detect signatures of non-thermal electrons and flows with velocities down to <50 km s$^{-1}$
- the ability to directly observe 2-D images rather than constructing them from a series of 1-D images
- photon-counting capability to provide millisecond or better time resolution, as well as achieving imaging times down to a few seconds.

The unique capabilities of these new detectors will allow us to study how magnetic energy is released in small scale magnetic reconnection events such as nanoflare heating in active regions and the quiet Sun.

An instrument optimized to be sensitive to the accurate detection of faint hot coronal plasma and plasma flows can make important contributions beyond the coronal heating problem if combined with other observations. These include but are not limited to:
- Identifying and characterizing the physics of interchange reconnections between open and closed fields which has been observed by AIA near coronal holes
- Characterizing the energetics of CME heating during eruptions which may represent a significant fraction of its energy budget (Akmal et al. 2001; Ciaravella et al. 2001; Rakowski et al. 2007; Lee et al. 2009; Landi et al. 2010)
- Investigate the density structures in solar flares which may influence the energetics of nonthermal electrons, using diagnostic line ratios in Fe XXI and Fe XXII (Phillips et al. 1996).

Decisive observations of the hot plasma associated with nanoflare models of coronal heating can be provided by new solar microcalorimeters. These measurements will cover the most important part of the coronal spectrum for searching for the nanoflare-related hot plasma and will characterize how much nanoflares can heat the corona both in active regions and the quiet Sun. Such measurements can also make great contributions to the study of the FIP effect by measuring relative abundances of a large number of important elements (e.g. Fe, Ca, Si, Mg, S, Ar, C, Ne, and O) at the same time. In addition, microcalorimeters will enable to study all of this as a function of time and space in each pixel simultaneously. NASA and JAXA have already collaborated successfully on microcalorimeters and would make ideal partners to apply them to solar observations.